\documentclass[%
 aip,
pra,
 amsmath,amssymb,
 reprint,
]{revtex4-2}

\usepackage{graphicx}% Include figure files
\usepackage{dcolumn}% Align table columns on decimal point
\usepackage{bm}% bold math
\usepackage[utf8]{inputenc}
\usepackage[T1]{fontenc}
\usepackage{mathptmx}
\usepackage{mathtools}
\usepackage{amsmath}
\usepackage{etoolbox}
\usepackage{braket}
\usepackage{color}
\usepackage{siunitx}
\usepackage[normalem]{ulem}

\usepackage[hidelinks]{hyperref}
\hypersetup{
  colorlinks   = true, %Colours links instead of ugly boxes
  urlcolor     = blue, %Colour for external hyperlinks
  linkcolor    = blue, %Colour of internal links
  citecolor   = blue %Colour of citations
}

\makeatletter
\def\@email#1#2{%
 \endgroup
 \patchcmd{\titleblock@produce}
  {\frontmatter@RRAPformat}
  {\frontmatter@RRAPformat{\produce@RRAP{\phantom{*}\!\!#1\href{mailto:#2}{#2}
  }}\frontmatter@RRAPformat}
  {}{}
}%
\makeatother

\newcommand{\myaffiliation}{\affiliation}%\address}

\newcommand{\IBEC}{\myaffiliation{IBEC -- Institute for Bioengineering of Catalonia, 08028 Barcelona, Spain}}

\newcommand{\Jlich}{\myaffiliation{Institute of Biological Information Processing (IBI-7), Forschungszentrum Jülich, 52425 Jülich, Germany}}

\newcommand{\ICREA}{\myaffiliation{ICREA -- Instituci\'{o} Catalana de Recerca i Estudis Avan{\c{c}}ats, 08010 Barcelona, Spain}}

\newcommand{\ICFO}
{\myaffiliation{ICFO -- Institut de Ci\`encies Fot\`oniques, The Barcelona Institute of Science and Technology, 08860 Castelldefels (Barcelona), Spain}}

\begin{document}

%\preprint{AIP/123-QED}

\title{Live magnetic observation of parahydrogen hyperpolarization dynamics}
% Force line breaks with \\

%^{*}The authors contributed equally.

\author{James Eills$^\dagger$}
\IBEC
\Jlich
\email[$\dagger$ Electronic mail: ]{j.eills@fz-juelich.de}

\author{Morgan W. Mitchell}
\ICFO
\ICREA

\author{Irene Marco Rius}
\IBEC

\author{Michael C.D. Tayler$^\ast$}
\ICFO
\email[$\ast$ Electronic mail: ]{michael.tayler@icfo.eu}

\date{\today. Accepted manuscript copy for arXiv.}

% At least three keywords are required at submission. Please provide three to five keywords, separated by the pipe symbol.
\keywords{Hyperpolarization $|$ NMR $|$ Parahydrogen $|$ Optical magnetometry $|$ Adiabaticity}

\begin{abstract}
\textbf{Significance statement:} Molecules can be prepared in a hyperpolarized state to enhance their nuclear magnetic resonance (NMR) signals by four to five orders of magnitude, enabling many applications in spectroscopy and imaging (MRI). 
But signal read-out typically requires a magnetic field pulse to perturb the system's quantum state, followed by signal detection.
This process destroys the hyperpolarized state, and provides low-frame-rate observations when tracking dynamic processes.
Here we use optical magnetometry to observe induced quantum state %transitions 
transformations of hyperpolarized molecules in real time.  This method offers an unprecedented view of hyperpolarization processes, revealing details analogous to switching from still-image film to video. Observation throughout the hyperpolarization process, even in the presence of time-dependent fields, opens possibilities for real-time control.\\

\noindent \textbf{Abstract:} Hyperpolarized nuclear spins in molecules exhibit high magnetization that is unachievable by classical polarization techniques, making them widely used as sensors in physics, chemistry, and medicine.
The state of a hyperpolarized material, however, is typically only studied indirectly and with partial destruction of magnetization, due to the nature of conventional detection by resonant-pickup nuclear magnetic resonance spectroscopy or imaging.
Here we establish atomic magnetometers with sub-\SI{}{\pico\tesla} sensitivity as an alternative modality to detect in real time the complex dynamics of hyperpolarized materials without disturbing or interrupting the magnetogenesis process.  
As an example of dynamics that are impossible to detect in real time by conventional means, we examine parahydrogen-induced \textsuperscript{1}H and \textsuperscript{13}C magnetization during adiabatic eigenbasis transformations at \si{\micro\tesla}-field avoided crossings.  Continuous but nondestructive magnetometry reveals previously unseen spin dynamics, fidelity limits, and magnetization backaction effects.  
As a second example, we apply magnetometry to observe the chemical-exchange-driven \textsuperscript{13}C hyperpolarization of [1--\textsuperscript{13}C]-pyruvate --- the most important spin tracer for clinical metabolic imaging.
The approach can be readily combined with other high-sensitivity magnetometers and is applicable to a broader range of general observation scenarios involving production, transport and systems interaction of hyperpolarized compounds.
\end{abstract}

\maketitle

\noindent Nuclear spin hyperpolarization is a family of non-equilibrium spin-manipulation methods, of interest across physics, chemistry, biomedicine and materials science.  Hyperpolarization provides a route to extremely strong spin magnetization, boosting NMR signals by orders of magnitude relative to equilibrium polarization by strong magnetic fields\cite{eills2023spin}. Hyperpolarized materials are used as particle-scattering targets\cite{ErhardSteffens2003,Niinikoski2020book}, clinical MRI probes of metabolism kinetics and metabolite biodistribution in vivo\cite{Kurhanewicz2019Neoplasia} and for characterizing structures of nanosheets, catalyst surface sites, and other low-abundance chemical samples\cite{rankin2019recent}.  Hyperpolarized materials also exhibit selective spin polarization, in which particular nuclei are strongly polarized while their neighbors remain unpolarized\cite{Hilty2022natureprotocols}.

\begin{figure*}
    \centering
    \includegraphics[width=\textwidth]{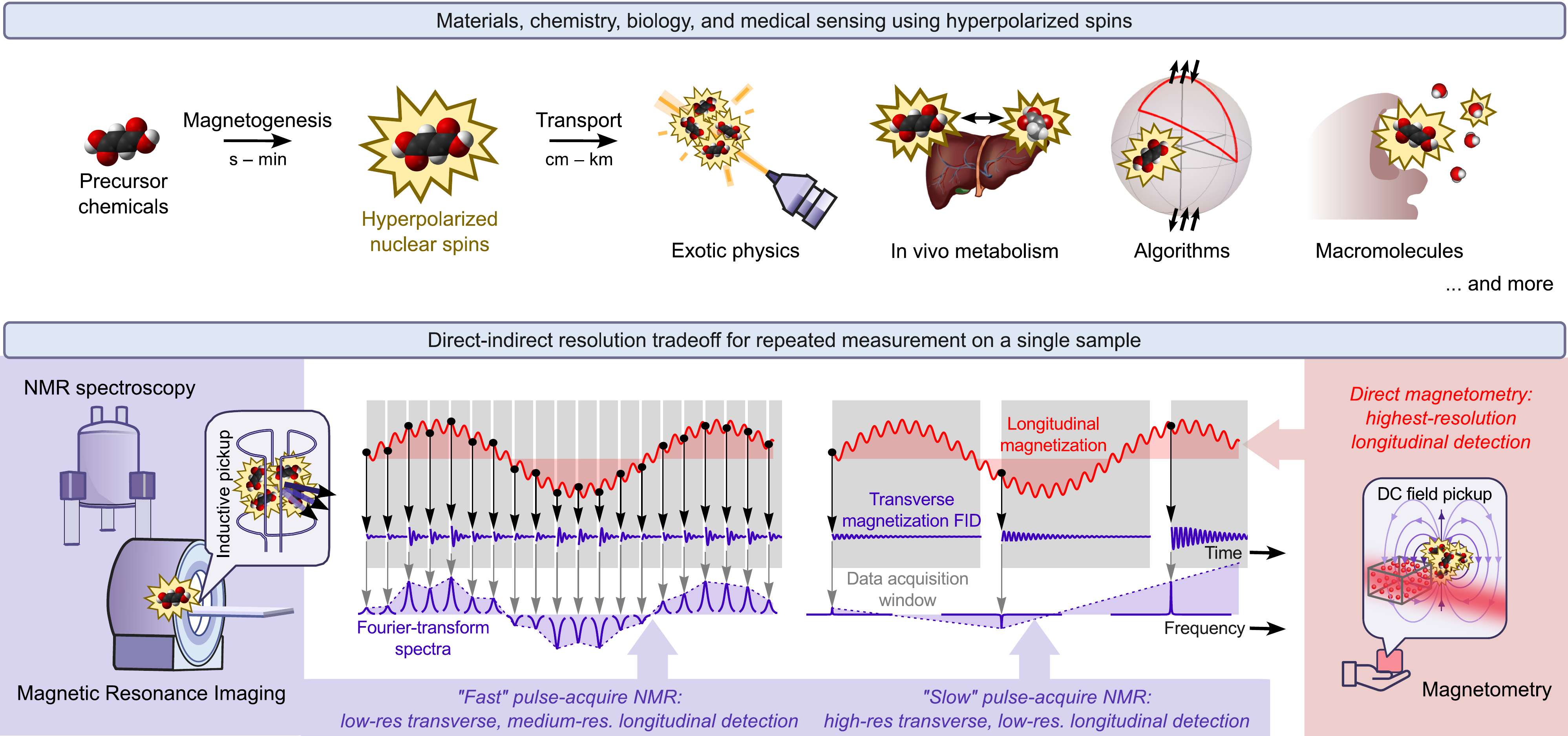}
    \caption[width=\textwidth]{Hyperpolarization boosts the NMR signals of chemicals and materials, enabling a range of applications from medical diagnostics to particle physics, but the process of exciting the signals for magnetic detection is slow, and is destructive to the hyperpolarized state. Additionally, detector saturation typically precludes simultaneous spin-driving and signal detection, necessitating indirect detection via multiple `hyperpolarize--observe' cycles.  Direct detection using optical magnetometry allows `live' hyperpolarization dynamics to be observed nondestructively, even in the presence of driving fields. 
    }
    \label{fig:fig1new}
\end{figure*}

Although powerful, capable of spectroscopically distinguishing nuclei by chemical shifts, conventional NMR imposes multiple limitations on the detection of hyperpolarized spins. Typically in NMR, one detects magnetization oscillations that are generated by tipping the magnetization off-axis by a radiofrequency (RF) pulse, then collecting the free-precession signal via RF inductive pickup.  This ``indirect'' approach requires intervention for signal excitation, (i) imposing a time-frequency bandwidth limitation determined by the pulse-and-acquire repetition rate (\autoref{fig:fig1new}), and (ii), consuming polarization in the process. It is common to use designed RF excitation pulse schemes to create or manipulate spin polarization\cite{eills2023spin,barskiy2016NMR}, but observing the intermediate dynamics with conventional NMR would require interrupting the pulse scheme for a detection period. The standard workaround is to repeat the experiment using multiple samples to create a time series, with each sample contributing one datapoint\cite{theis2011parahydrogen,theis2012zero,Blanchard2020JMR,Savukov2020JMR,blanchard2021towards,van2022relayed,put2023detection,dagys2022hyperpolarization,pravdivtsev2021coherent,eills2019polarization,rodin2021constant,colell2017generalizing,tanner2019selective,myers2024zero}, but this comes at the cost of time, material, and the introduction of random error. 
Moreover such an approach cannot be used with one-time-hyperpolarized samples such as those destined for in vivo sensing\cite{nelson2013metabolic}, and those exhibiting complex or chaotic behavior\cite{Marion2008ChemPhysChem}.

To overcome these limits and substantially widen the opportunities for sensing using hyperpolarized spins, an excitation-free detector is required, to sample the magnetization directly. In this work, we demonstrate that an optically pumped magnetometer (OPM) as a suitable direct detector. As its sensing medium, the OPM employs the valence electrons of alkali-metal atoms (e.g., Rb) in the vapor phase\cite{BudkerOpticalMagnetometryBook}.  Owing to the high magnetic moment, coherence time, and number of polarized spins in the vapor, OPMs can precisely measure the fT-pT fields produced by hyperpolarized\cite{cohentannoudji1969detection,theis2011parahydrogen,theis2012zero,barskiy2019zero,sheberstov2021photochemically,eills2023enzymatic,put2023detection} or thermally-polarized NMR ensembles\cite{Savukov2005PRL,Tayler2017RSI,Bodenstedt2021natcomm}, over a range of frequencies from DC to kHz\cite{Ganssle2014ANIE,mouloudakis2023real,Blanchard2020JMR}. 
The OPM's ability to non-destructively observe spin dynamics in real time opens a unique window into complex hyperpolarization processes. As we show, high-resolution magnetization tracking via OPMs can elucidate transfer mechanisms, allow speed optimization of sequences, and potentially enable feedback control of hyperpolarization.

These features are illustrated using parahydrogen-induced polarization (PHIP),\cite{bowers1986transformation,bowers1987parahydrogen,eisenschmid1987para} a particularly powerful hyperpolarization technique for liquid samples that can polarize nuclear spins to nearly 100\% in a range of systems, from dilute analytes to highly concentrated solutes\cite{Nagel2023}. PHIP operates by cooling H\textsubscript{2} gas into a spin-singlet state, transferring the low-entropy spin order to the nuclei of a target molecule, and converting this spin order to vector-polarized form using control fields, which can range in strength from sub-\si{\micro\tesla} \cite{johannesson2004transfer,  cavallari2015effects, eills2019polarization,  rodin2021constant, lindale2019unveiling, pravdivtsev2021coherent, marshall2023radio} to T \cite{goldman2006design,schmidt2017liquid,knecht2018mechanism,eills2017singlet}. 
Specifically, we use the OPM to demonstrate longitudinal magnetization detection during PHIP magnetogenesis: (1) nondestructively, (2) beyond the resolution-bandwidth limit of conventional NMR, where the sampling rate is much faster than the Larmor frequency, and (3) during ongoing adiabatic transitions.  Features unveiled include the sensitivity of PHIP processes to spin-spin couplings, fidelity of repeated magnetizing and demagnetizing field cycles, and backaction effects of distant dipolar fields\cite{edzes1990nuclear,levitt1996demagnetization,dagys2024robust}.

\section*{Results}

In this section we highlight the unique features of direct sampling through the ability to unveil information on magnetization dynamics in PHIP and related hyperpolarization processes. A more general interpretation of the capabilities follows in the Discussion (\autoref{sec:discussion}).

\subsection{Observing Magnetogenesis during Adiabatic Eigenbasis Transformations}

\begin{figure*}[t]
    \centering
    \includegraphics[width=\textwidth]{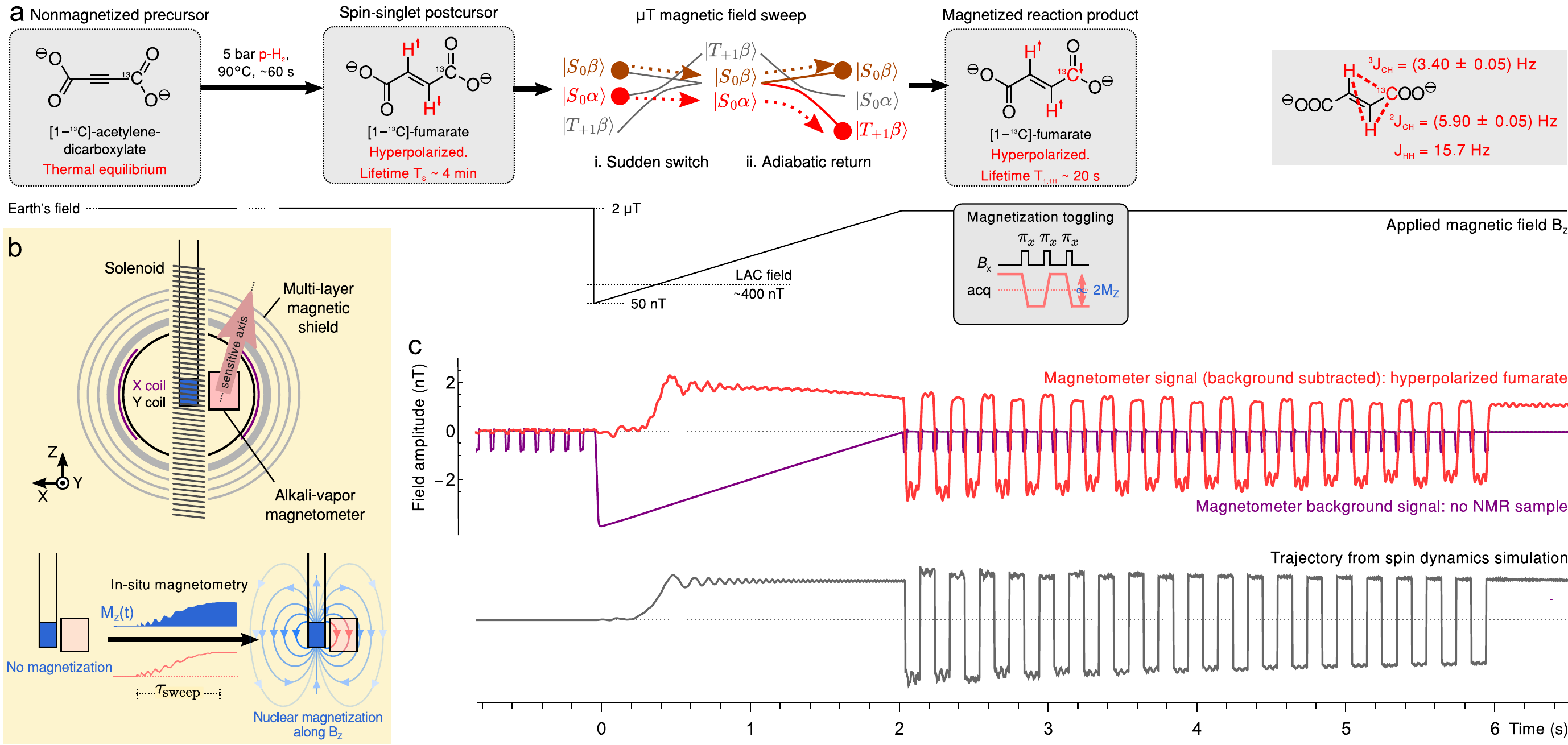}
    \caption{Direct observation of magnetogenesis in nuclear-spin-hyperpolarized molecules. 
    (a) Hyperpolarized fumarate is formed by reducing [1--\textsuperscript{13}C]-acetylenedicarboxylic acid with para enriched H\textsubscript{2}, and a \si{\micro\tesla} magnetic field sweep (either linear or constant adiabaticity) is applied for scalar-to-vector order conversion. After magnetogenesis, the nuclear spin vector order may be toggled up and down along the $z$ axis by a series of $\pi$ pulses, or excited with a pulse of tip angle $\theta\neq\pi$ to produce oscillating transverse magnetization.
    (b) A solenoid coil supplies the sweep field, and an alkali-vapor OPM (in the solenoid's exterior rather than interior field to maximize operating range) is used for detection. Both are located inside a multi-layer magnetic shield.
    The OPM's sensitive axis is tilted $\sim$\SI{30}{\degree} away from the solenoid axis to be sensitive to both longitudinal ($z$) and transverse ($x$ and $y$) fields produced by the sample.
    (c) Typical OPM pickup signals during a 50\,nT to \SI{2}{\micro\tesla} field sweep: (purple) stray field of the solenoid field during the sweep; (red) stray field of hyperpolarized [1--\textsuperscript{13}C]-fumarate (1\,mL, 60\,mM).
    Magnetization toggling before and after the field sweep is performed using composite $\pi$-pulses (90\textsubscript{{X}}180\textsubscript{{Y}}90\textsubscript{{X}}) selective to \textsuperscript{1}H, and spaced at 0.1\,s intervals.  
    After stopping at 6\,s the magnetization persists as a stationary state in the solenoid field.  
    The simplified state energy diagrams show population exchange during the adiabatic sweep, leading to a magnetized state. To the right, the [1--\textsuperscript{13}C]-fumarate $J$-coupling parameters are shown. These were used, within the specified errors, for all subsequent simulations. 
    % \rtext{Figure changes: 
    % (1) increase size of j coupling diagram, right align with the figure edge. 
    % (2) increase size of adiabatic conversion diagram, darker gray
    % (3) generally increase font size, make as uniform as possible}
    }
    \label{fig:fig1}
\end{figure*}

As a first example, we illustrate direct magnetometry of hyperpolarized fumarate, a metabolically relevant compound that is prepared by reaction of parahydrogen with an unsaturated precursor, acetylenedicarboxylate (\autoref{fig:fig1}a).   The hydrogenation product is initially nonmagnetic due to the para spin state of the \textsuperscript{1}H pair, but can be converted to net magnetization by adiabatic cycling in sub-\si{\micro\tesla} fields.  Many variants of this procedure exist\cite{johannesson2004transfer,cavallari2015effects,eills2019polarization,joalland2019pulse,rodin2021constant} -- the one shown involves a sudden switching of the background field down to a few nT and then slowly increasing the field linearly up to \SI{2}{\micro\tesla}, passing through the avoided state crossing near \SI{400}{\nano\tesla}. 

In the ideal adiabatic case, three of the molecule's eight nuclear-spin angular momentum states are involved.  Initially, molecules are equally distributed within two of the high-field eigenstates, $\ket{S_0\alpha}$ and $\ket{S_0\beta}$, and the state corresponds to \textsuperscript{1}H-only spin order.  Here the first label refers to the \textsuperscript{1}H singlet state $\ket{S_0} \equiv \ket{0,0}$ with zero total angular momentum and the second label refers to the state of \textsuperscript{13}C, with total angular momentum $\hbar/2$ and projection $+\hbar/2$ ($\ket{\alpha} \equiv \ket{1/2,\,+1/2}$) or $-\hbar/2$ ($\ket{\beta} \equiv \ket{1/2,\,-1/2}$). 
Sweeping the magnetic field from 0 to 2\,µT causes population exchange between $\ket{S_0\alpha}$ and the third state $\ket{T_{+1}\beta} \equiv \ket{\alpha\alpha\beta}$. $\ket{S_0\beta}$ is not mixed with other states and retains its population throughout.  The final state is equal populations of $\ket{T_{+1}\beta}$ and $\ket{S_0\beta}$ which corresponds to equal \textsuperscript{1}H and \textsuperscript{13}C polarization, with different sign. Hence, a single sweep leads to both \textsuperscript{1}H and \textsuperscript{13}C magnetization, with relative amplitude $(+\gamma_{\rm H})$:$(-\gamma_{\rm C})$, or approximately 4:$-1$.

Magnetometry and field cycling involves remarkably simple hardware comprising the OPM, a solenoid coil, a low-voltage programmable waveform generator and a passive magnetic shield, (\autoref{fig:fig1}b and Methods section).  From a geometric standpoint, the solenoid coil is highly accommodating towards integrating the OPM; in an ideal scenario, the enclosed sample encounters a uniform magnetic field due to the solenoid, while the external OPM experiences no field.\cite{yashchuk2004hyperpolarized}  Consequently, the impact of the field cycle on the magnetometer's dynamic range is zero.  Real-world conditions deviate only slightly from this ideal; the stray field generated by the solenoid is at least 500 times weaker than the internal field, so the OPM response is linear across several tens of \si{\micro\tesla} in terms of the background field at the hyperpolarized sample. 

To measure the field produced by the sample without background field contributions or other slow drifts in the OPM, the sign of the nuclear magnetization is repeatedly toggled via a series of high-fidelity composite \SI{180}{\degree} pulses (\autoref{fig:fig1}c).  When toggling is applied before the field cycle, no change is seen in the overall signal -- confirming the initial hyperpolarized state is nonmagnetic.  
In contrast, after magnetization, a clear alternating pattern is seen where the peak-to-peak amplitude ($\sim$\SI{4}{\nano\tesla}) equals twice the field produced by the sample.  The \textsuperscript{13}C polarization for these experiments was measured separately in a high-field NMR system to be approximately 40\%.
Spin-selective toggle pulses allow magnetization from different nuclei to be inverted, such as in \autoref{fig:fig1}c where the 90\textsubscript{{X}}180\textsubscript{{Y}}90\textsubscript{{X}} composite $\pi$ pulses are applied to give a net rotation of \SI{180}{\degree} on \textsuperscript{1}H; these are mostly selective to \textsuperscript{1}H magnetization, since they yield a 22.5\textsubscript{{X}}45\textsubscript{{Y}}22.5\textsubscript{{X}} or around \SI{63}{\degree} net rotation of the \textsuperscript{13}C magnetization, which quickly loses coherence after a few toggling repetitions.  This reveals that the observed magnetic field is strongly weighted towards the \textsuperscript{1}H magnetization -- as expected from the $\sim$$3.97\times$ higher gyromagnetic ratio of \textsuperscript{1}H -- even though the field sweep should polarize \textsuperscript{1}H and \textsuperscript{13}C equally (but in opposite directions).  
A small \textsuperscript{13}C  magnetization is evident from the coherent oscillations superposed upon the toggled \textsuperscript{1}H signal, at the \textsuperscript{13}C Larmor frequency \SI{2}{\micro\tesla}\,$\times \gamma_{\rm C} = 2\pi \times \SI{21.5}{\hertz}$.

\begin{figure*}
    \centering
        \includegraphics[width=0.9\textwidth]{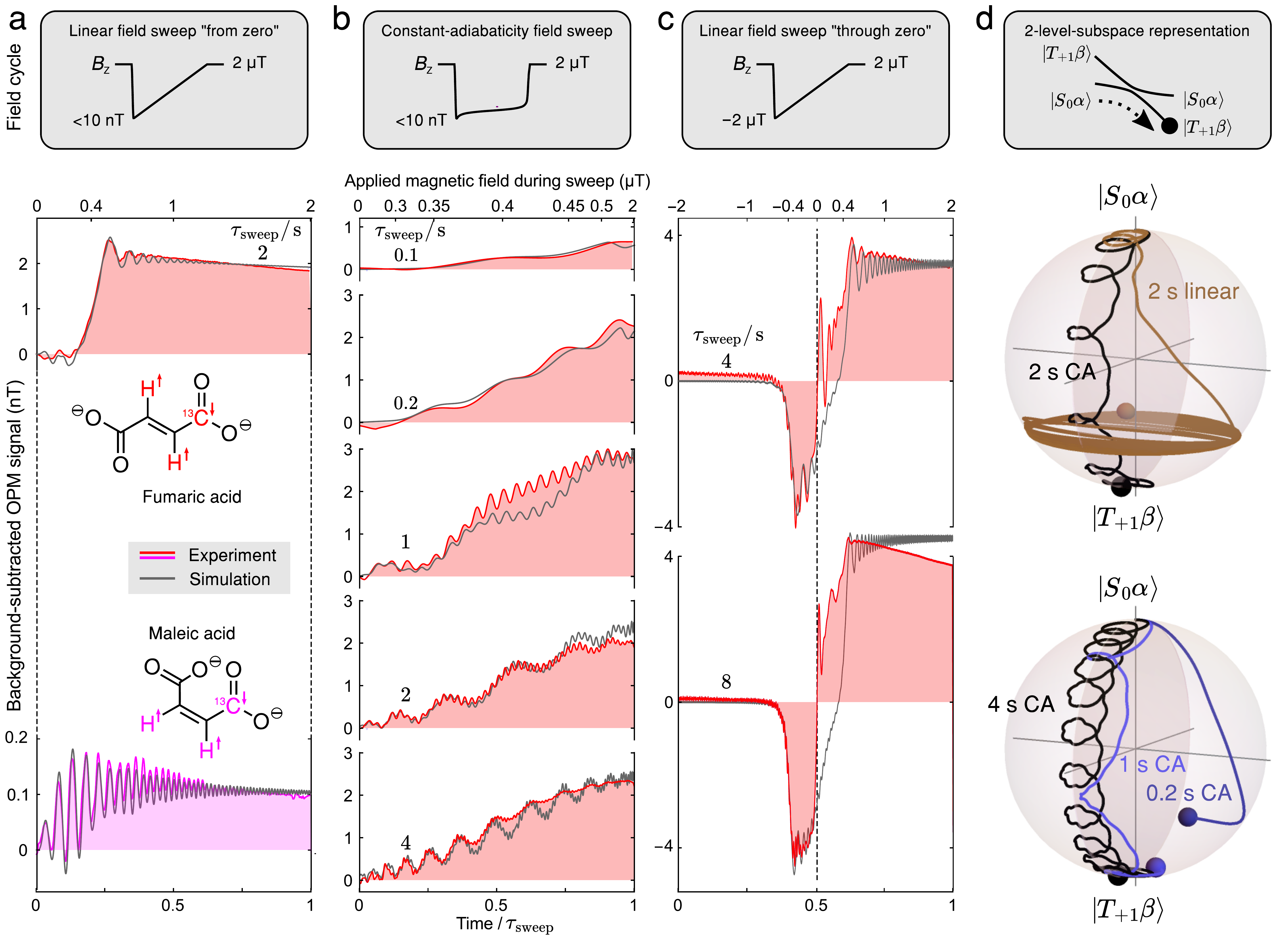}
    \caption{Experimental (color) and simulated (gray) magnetic field profiles for different magnetogenesis cases. (a) OPM signals detected during a linear 0-2\,$\mu$T field sweep for the \textit{trans} and \textit{cis} isomers of [1--\textsuperscript{13}C]-butenedicarboxylate. The applied field sweep is the same for both molecules, but vastly different magnetogenesis dynamics occur due to different $J$-couplings 
    (\textit{trans}
    \textsuperscript{3}$J$\textsubscript{HH} $=$ \SI{15.7}{Hz}, 
    \textsuperscript{2}$J$\textsubscript{CH} $=$ \SI{5.94}{Hz}, 
    \textsuperscript{3}$J$\textsubscript{CH} $=$ \SI{3.4}{Hz}; 
    \textit{cis} 
    \textsuperscript{3}$J$\textsubscript{HH} $=$ \SI{12.2}{Hz}, 
    \textsuperscript{2}$J$\textsubscript{CH} $=$ \SI{13.4}{Hz}, 
    \textsuperscript{3}$J$\textsubscript{CH} $=$ \SI{2.6}{Hz}).
    The weaker maleate (\textit{cis}) signal arises from its lower concentration, due to the slower chemical reaction. 
    (b) OPM signals detected during constant-adiabaticity (CA) field sweeps of differing total duration. These profiles are smoother than for uniform field sweeps, although additional low- and high-frequency oscillations are apparent; see panel (d).
    (c) OPM signals under a field inversion from $-2$ to $+2$\,$\mu$T. Vastly different profiles from those in (a) and (b) are observed, with an initial negative magnetization build up followed by a rapid inflection yielding positive magnetization. 
    Simulations represent the total magnetization produced by the 3 spins, vertically scaled to fit the experimental data; the same vertical scaling is applied to each set of panels in (b) and (c).  In (a), phenomenological relaxation is included in the simulated profiles, where $T_1$ and $T_2$ are given in the Supporting Information. 
    (d) 3D (Bloch sphere) representation of magnetogenesis trajectories for [1--\textsuperscript{13}C]-fumarate during the 0 to $+2$\,$\mu$T sweeps.  Spin states $\ket{S_0\alpha}$ and $\ket{T_{+1}\beta}$ as defined in the main text form a near-isolated two-level subspace that undergoes population inversion during sweeps through the anticrossing field.  Magnetization is proportional to the vertical distance from the north pole.
    CA sweeps yield higher-fidelity inversion than the linear sweep, and the beat features are clearly visualized as ``cycloid arcs''.
    }
    \label{fig:fig2}
\end{figure*}

High-resolution magnetization tracking reveals prominent features between \SI{0}{\second} to \SI{2}{\second} where during the linear ramp the magnetization rapidly increases around the \SI{400}{nT} anticrossing field.  Here the spin eigenbasis changes most rapidly with respect to magnetic field, with the exact shape of the profile depending on the scalar ($J$) couplings between the spins.  A shape difference is illustrated in \autoref{fig:fig2}a between fumarate and its \textit{cis} isomer, maleate.  The center field of the crossing depends on the \textsuperscript{1}H-\textsuperscript{1}H coupling $J_\text{HH}$, and the state energy difference at the center field mostly depends on $\Delta J\textsubscript{CH} = \textsuperscript{2}J_\text{CH}-\textsuperscript{3}J_\text{CH}$,\cite{eills2019polarization} where the superscript denotes the number of chemical bonds between nuclei.  The large oscillations in the \textit{cis} isomer ([1-\textsuperscript{13}C]-maleate) profile occur due to a larger value of $\Delta J_{\rm CH}$.  Values for $J_{\rm HH}$ and $J_{\rm CH}$ are provided in \autoref{fig:fig1}.

Direct magnetization tracking immediately reveals that the speed of magnetogenesis in the linear-ramp case is limited by adiabaticity requirements, particularly around the level anticrossing region.  
Enhanced or `constant adiabaticity' ramp profiles can be calculated to pass quickly through regions of the ramp where the eigenbasis does not change with field, but slowly through the important anticrossing region\cite{rodin2019constant}.
The OPM signal recorded during such a profile (see \autoref{fig:fig2}b) shows a more uniform magnetization build-up rate compared to the linear profile.  
The increased average speed also helps to reduce relaxation losses, though these are not evident for fumarate because the ramp duration is much shorter than the \textsuperscript{1}H magnetization decay time constant $T_1\sim$ \SI{33}{s}, see \autoref{fig:fig4}.

\begin{figure*}
    \centering
    \includegraphics[width=\textwidth]{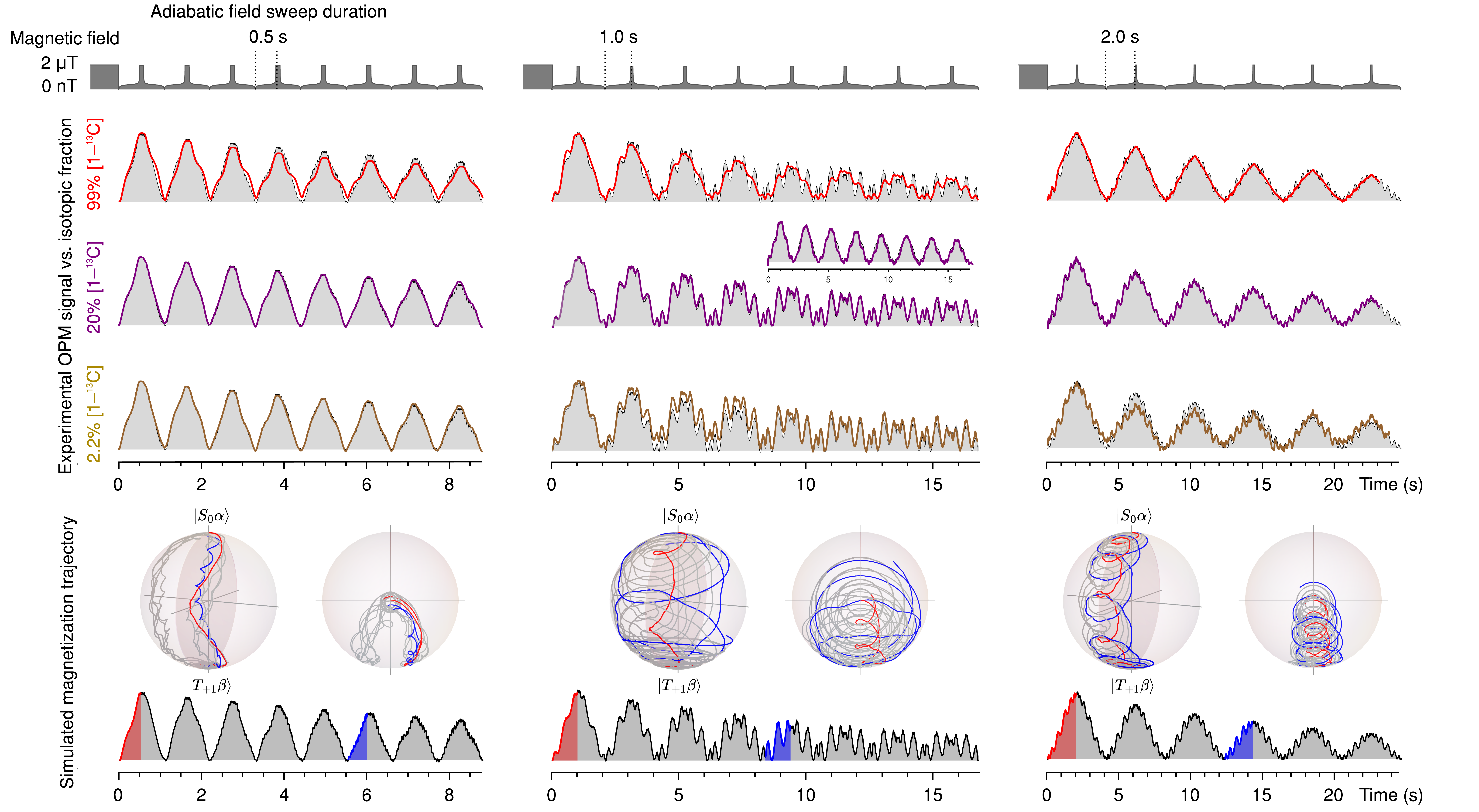}
    \caption{Reversibility of the magnetogenesis process.  Magnetizing (\SI{0}{\micro\tesla} to \SI{2}{\micro\tesla}) and demagnetizing (\SI{2}{\micro\tesla} to \SI{0}{\micro\tesla}) sweeps are performed repeatedly with a \SI{100}{\milli\second} pause at \SI{2}{\micro\tesla} with continuous detection throughout.
    Top: Applied magnetic field profiles and a simulation of the total nuclear magnetization for [1--\textsuperscript{13}C]-fumarate for three different sweep durations (\SI{0.5}{\second}, \SI{1}{\second}, \SI{2}{s}).
    Center: Experimental data showing the magnetic field produced by samples of [1--\textsuperscript{13}C]-fumarate subjected to the field sweeps, at different levels of \textsuperscript{13}C isotopic enrichment (2.2\%, 20\%, 99\%); the gray filled region denotes simulation. 
    Bottom: Simulated magnetization trajectories for sweeps between \SI{0}{\micro\tesla} and \SI{1.986}{\micro\tesla} in the absence of additional fields, such as the sample's internal field.  The sphere diagrams show selected regions of the trajectory in the $\{\ket{S_0\alpha}$, $\ket{T_{+1}\beta}\}$ two-level subspace at the beginning and center of the profiles, in red and blue, respectively.
    Fitted exponential decay time constants for the simulated profiles are between \SI{16}{\second} and \SI{22}{\second} (see Supporting Information). The inset for the 20\% isotopic labeling, \SI{1}{\second} result shows that a different magnetization trajectory (in both experiment and simulation) is observed when the $B_z$ field during the 100\,ms pauses is set to 2.05\,µT.} %different experimental and simulated curves for repeated sweeps between \SI{0}{\micro\tesla} and \SI{2.05}{\micro\tesla}. 
    \label{fig:fig3}
\end{figure*}

For the constant-adiabaticity case, our approach reveals expected\cite{Ivanov2022JMRanalyticaladiabatic} but previously unseen oscillations in the magnetization build-up curve, which appear to have a near-constant frequency across most of the sweep profile.  The slow ones indicate non-adiabatic behavior and are known as ``cycloid arcs'' when the state of the 2-level system is represented as a trajectory on a Bloch sphere (see \autoref{fig:fig2}d). 
Here, the cycloid arcs are mainly due to small coherences between states $\ket{S_0\alpha}$ and $\ket{T_{+1}\beta}$ that are introduced during the initial field switch to \SI{0}{\micro\tesla}. 
The faster oscillations correspond to a weakly-populated magnetic state that precesses around the applied $B_z$ field (which for the majority of the sweep is $\sim$400\,nT). This state is formed by the toggling pulses prior to the field sweep, which convert a small fraction of the population imbalance between \textsuperscript{1}H $\ket{S_0}$ and $\ket{T_0}$ states into coherences.
Simulations show that these oscillations are predominantly sensed in the $x$ and $y$ axes, and are $\sim$$10\times$ smaller if the toggling pulses are omitted.
These and other features are visible even at low concentrations of the hyperpolarized product, including natural \textsuperscript{13}C isotopic abundance, due to the low total noise ($\sim1$ pT\textsubscript{RMS}, 0--100 Hz) of the OPM.  
The exact shape of the peak/trough features can vary with physical conditions such as pH and temperature, which can alter, e.g., the asymmetry of the heteronuclear $J$\nobreakdash-couplings. 
Real-time observation can be used to understand these dependencies, but also to minimize their nuisance effects: the magnetization process could be stopped at a beat's peak, for example, thus optimizing the polarization yield at any pH and temperature. In complex, multi-spin systems, these features could also be used to understand magnetogenesis behavior that simplistic but more intuitive models (involving only a few spins) do not account for.  

An alternative to the 0 to $+$\SI{2}{\micro\tesla} field sweeps for singlet-to-magnetization conversion are bipolar field sweeps that pass through zero field\cite{eills2019polarization,joalland2019pulse,rodin2021constant}, e.g., $-$\SI{2}{\micro\tesla} to $+$\SI{2}{\micro\tesla}. Magnetogenesis during a bipolar field sweep is shown in \autoref{fig:fig2}c.  The net magnetization remains close to zero until the system first passes through an anticrossing field, then builds up in one direction, before changing direction after subsequent anticrossing fields.
Simulations agree with these inflections of the magnetization, but in the experimental data, additional peaks of $\sim$2\,nT magnitude can be seen.  The latter may be caused by small ($\sim$10\,nT) transverse fields that diabatically excite $\ket{T_{+1}}\leftrightarrow\ket{T_0}$ coherences when passing through the zero field crossing; thus, providing for free further information that would require a highly dedicated effort with conventional NMR sampling.

\subsection{Fidelity of Demagnetization and Remagnetization}

The adiabatic magnetogenesis process is reversible, meaning the sample can be reverted back to a nonmagnetic state of singlet order by reversing the applied field profile.  The magnetization trajectories across multiple magnetizing (\SI{0}{\micro\tesla} $\rightarrow$ \SI{2}{\micro\tesla}) and demagnetizing (\SI{2}{\micro\tesla} $\rightarrow$ \SI{0}{\micro\tesla}) field-sweep cycles can be measured and features of the curves analyzed to infer the fidelity of the process. For instance, in the perfect adiabatic case: (1) at the end of each field-sweep cycle the sample magnetization is zero; (2) a consistent magnetization is sustained across multiple cycles, limited only by the spin relaxation time constant.  In the fumarate system both of these features are present, indicating only minor losses (see \autoref{fig:fig3}, for sweep times 0.5\,s, 1\,s and 2\,s).  Even after 6 to 8 cycles the peak magnetic field exceeds half of the peak value during the first cycle.  This corresponds to a singlet-to-singlet fidelity of roughly 0.9 per cycle, mostly dominated by non-adiabatic losses in the short-sweep-duration case.  Long-term, high-resolution measurements such as these can be extended for up to minutes at essentially no cost, whereas for conventional pulse-acquire NMR approaches this would be prohibitive.

A clear contrast can be seen between the magnetization profiles for sweeps of 0.5\,vs.\,2\,s duration. The former appear smooth, whereas the latter show faster ($\sim$\SI{3}{\hertz}) oscillations on top, which are cycloid arcs due to coherences between states $\ket{S_0\alpha}$ and $\ket{T_{+1}\beta}$. The 1\,s sweep profile shows cycloid arc oscillations that appear to amplify over time, and eventually dominate the magnetization profile. This interesting behaviour is predicted by simulations, and is due to the sweep repetition rate coinciding with an overtone of the cycloid arc frequency, producing a resonance match. Simulations show that this effect may be suppressed by changing the field to avoid resonance. This can be verified both experimentally and by simulation; the inset in \autoref{fig:fig3} shows a smoother magnetization trajectory when the field during the \SI{100}{\milli\second} pause is changed from \SI{1.986}{\micro\tesla} to \SI{2.050}{\micro\tesla}.

\subsection{Quantifying the Sample's Internal Magnetic Field and Backaction on Magnetogenesis}

\begin{figure*}
    \centering
    \includegraphics[width=0.95\textwidth]{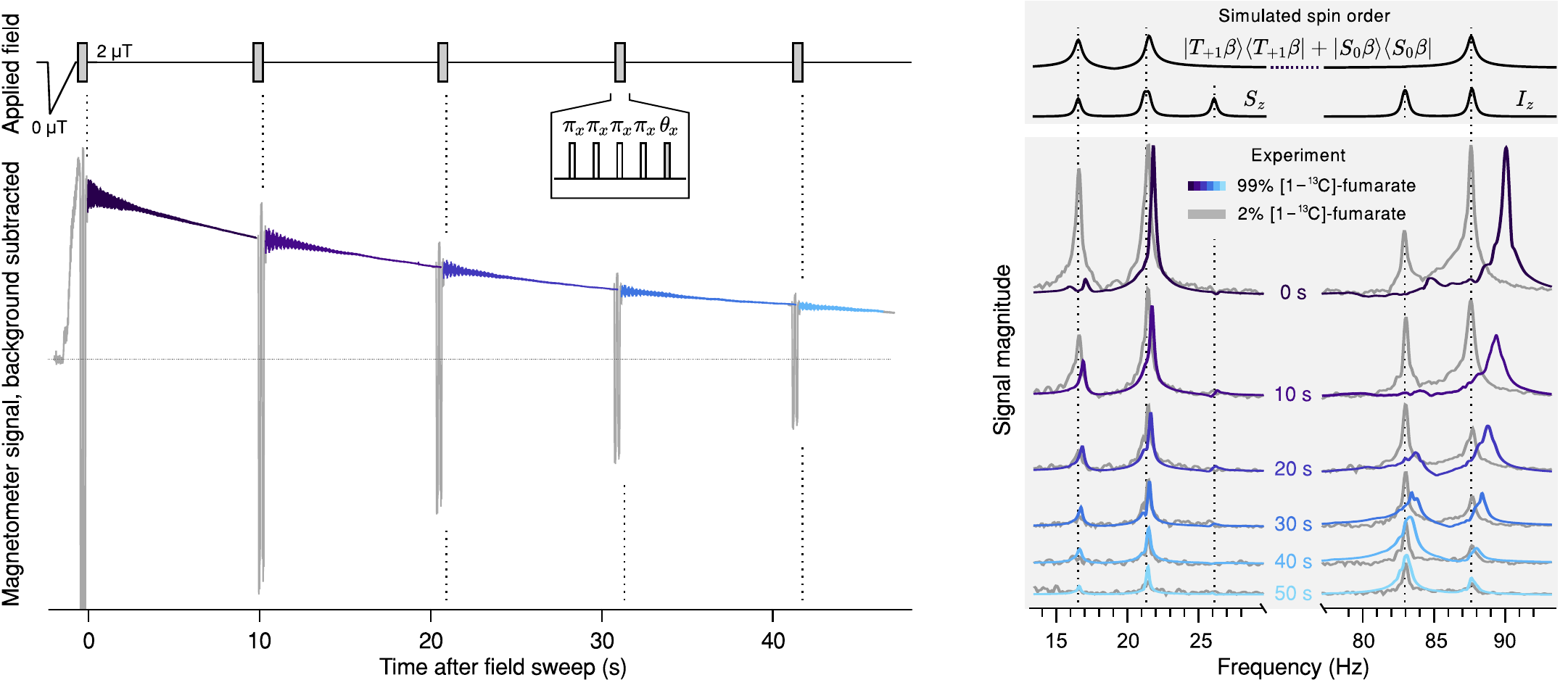}
    \caption{
    Time dependence of the \textsuperscript{1}H and \textsuperscript{13}C resonance frequencies in [1--\textsuperscript{13}C]-fumarate after magnetization following a 0 to \SI{2}{\micro\tesla} field inversion. Left: The pulse sequence employs short periods of magnetization toggling followed by a small-tip-angle pulse ($\theta\sim$\SI{10}{\degree}) to excite a free-precession NMR signals.
    Right: Fourier transforms of the individual signal transients for \textsuperscript{13}C-enriched and natural-abundance [1--\textsuperscript{13}C]-fumarate, showing the \textsuperscript{13}C and \textsuperscript{1}H signal regions separately at 13--29\,Hz and 78--93\,Hz. For natural-abundance [1--\textsuperscript{13}C]-fumarate the signals are fixed over time at the predicted resonance frequencies, but for the \textsuperscript{13}C-enriched sample the signals exhibit an exponentially decaying drift over time, arising from the sample's internal field. Colored signals are vertically scaled by a factor 0.01 relative to the gray signals. %\rtext{Consider stretching the left-hand plot so that the overall figure spans the full two-column width.}
    }
    \label{fig:fig4}
\end{figure*}

Nowadays, just as in the beginnings of pulsed NMR where RF radiation-damping effects were being noticed in concentrated liquids\cite{edzes1990nuclear,levitt1996demagnetization}, (e.g., water), practitioners of hyperpolarization are quickly realizing the importance of a sample's own magnetic moment or ``distant dipolar field'' (DDF) that behaves as an additional temporal, spatially nonuniform and possible chaotic\cite{Lin2000ScienceCrushedMagnetization} magnetic field experienced by the spins.
The DDF may interfere with the magnetization process itself\cite{dagys2024robust} when comparable in size to the applied magnetic fields.  Evidence for these effects can be observed for high ($>$100\,mM) concentrations of the hyperpolarized [1--\textsuperscript{13}C]-fumarate isotopolog as shown in \autoref{fig:fig3}, which compares the magnetization trajectories between samples of different labeling (2.2\%, 20\% and 99\% [1-\textsuperscript{13}C]).
For dilute samples (e.g., 2.2\% and 20\% [1--\textsuperscript{13}C]), the experimental data and simulated trajectories agree well, but beyond this (e.g., 99\% [1--\textsuperscript{13}C]), the number of magnetized molecules is sufficiently high for dipolar-field effects to be non-negligible.  The build-up curves are noticeably smoother, with the cycloid arc features being less pronounced.  This is attributed to destructive interference caused by the inhomogeneity of the DDF, and can be confirmed by comparing with numerically simulated profiles (inhomogeneity around $10\%$). This interference has been shown in other work to strongly limit polarization in PHIP experiments\cite{dagys2024robust}, but could only be studied after -- not during -- magnetogenesis. Direct tracking is advantageous because it allows one to understand and quantify DDF effects in real-time, which is subject to investigation in ongoing studies.

The absolute value of the sample-generated magnetic field detected at the OPM is not trivial to interpret quantitatively because it depends on the sample/detector shape, geometry and standoff distance, among other variables.  However, the OPM is capable of directly and rapidly quantifying the DDF in situ by pulse-acquire detection.  This approach measures Larmor frequency shifts immediately after the magnetizing field sweep by holding the sample at a low field (e.g., \SI{2}{\micro\tesla}) and periodically applying small ($\sim$\SI{10}{\degree})-tip-angle pulses to rotate the magnetization into the $x$/$y$ plane, as shown in \autoref{fig:fig4}.  The \textsuperscript{1}H precession frequency peak in highly magnetized fumarate (99\% \textsuperscript{13}C) is initially shifted by +3\,Hz, corresponding to a +70\,nT internal field experienced by the spins.  A smaller shift, scaled by the \textsuperscript{1}H/\textsuperscript{13}C gyromagnetic ratios, is also observed for the \textsuperscript{13}C in the same direction.  Both frequency shifts decay with a time constant of 30 - 33\,s, matching the 33\,s $T_1$ decay time constant of the DC magnetization (from the time-domain magnetometer signal).  Repeating this experiment at natural \textsuperscript{13}C isotopic abundance ($\sim$2\% for fumarate due to symmetry) yields no observable frequency shift, due to the significantly lower overall magnetization. 

If desired, other spectral features can be used to quantify these high magnetization effects further.  For instance, in the small-tip-angle approximation, the relative intensities of $J$-coupling multiplet lines correspond quantitatively to the relative population difference across spin states in the high-field eigenbasis, and therefore the absolute polarization of each spin.\cite{Vuichoud2015JMR260} 

\subsection{Observing Magnetogenesis during Chemical Exchange at Fixed Field}

We employed real-time OPM detection to observe hyperpolarization of the clinical metabolite [1--\textsuperscript{13}C]-pyruvate, to illustrate that the method is close to applications in patient MRI\cite{nelson2013metabolic,Kurhanewicz2019Neoplasia,Deen2023Radiology}, as well as that the OPM detection approach is applicable to different hyperpolarization methods. In this case, the \textsuperscript{1}H and \textsuperscript{13}C spins are polarized by catalytic exchange with parahydrogen, and because the pyruvate molecule is not chemically modified these can be depolarized and repolarized many times.  A simple version of the experiment (termed SABRE-SHEATH\cite{Theis2015sabresheath}) takes place with a solution of pyruvate and an iridium catalyst, through which parahydrogen is bubbled while being held in a sub-\si{\micro\tesla} field\cite{iali2019hyperpolarising} adjacent to the OPM (see Methods section).

\begin{figure*}
    \centering
    \includegraphics[width=\textwidth]{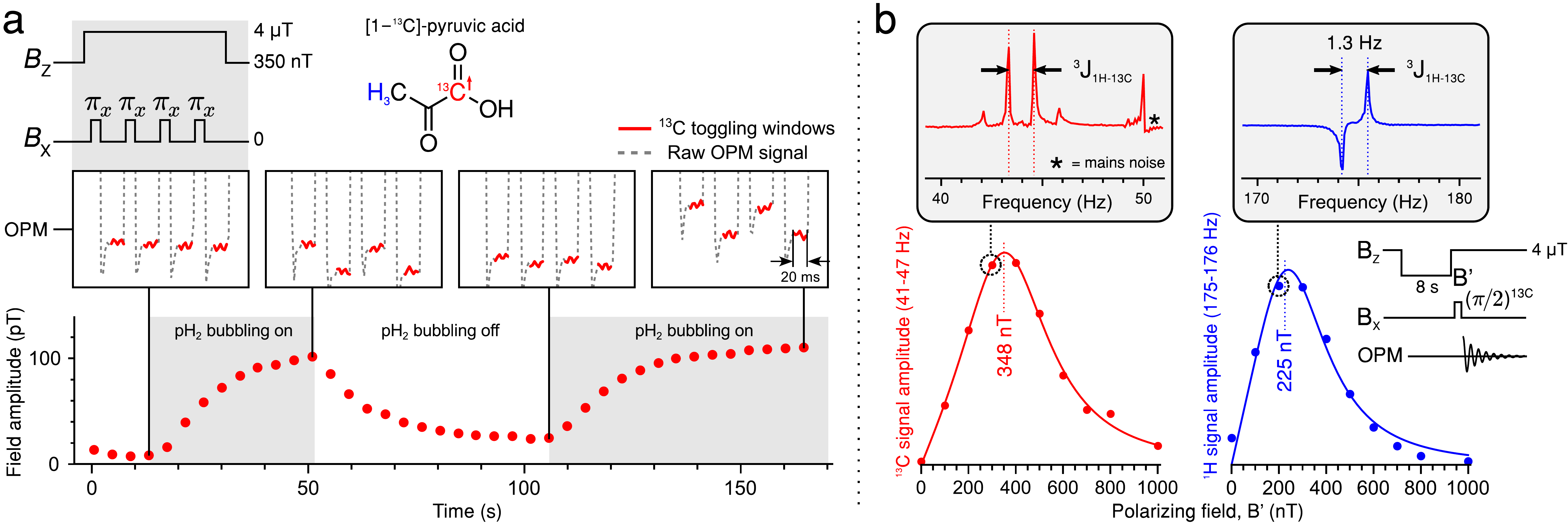}
    \caption{Magnetometry during SABRE-SHEATH polarization. (a) An experiment to probe \textsuperscript{13}C magnetization accumulation in 1\,mL, 60\,mM [1-\textsuperscript{13}C]-pyruvate at 350\,nT by employing brief intervals of \textsuperscript{13}C toggling at \SI{4}{\micro\tesla}. Parahydrogen bubbling was switched on and off to alternate between periods of polarization buildup and decay.  Plot markers represent the average measured magnetization during a toggling interval, and the raw magnetometer signal during the toggling interval is shown above (in red) for selected data points (mean signal difference between spin-up and spin-down toggling regions). (b) Optimization of the SABRE-SHEATH field dependence for the polarization of both \textsuperscript{1}H and \textsuperscript{13}C in [1-\textsuperscript{13}C]-pyruvate. Parahydrogen was bubbled continuously, and the magnetic field was fixed for an 8\,s polarization period at the field $B'$ followed by NMR signal excitation and detection at 4\,$\mu$T. In the upper plots representative \textsuperscript{13}C and \textsuperscript{1}H spectra are shown.
    In the lower plots, the integral of the absolute-mode NMR peaks in the specified frequency range is plotted against the polarizing field $B'$. The inset shows the pulse sequence used to collect the \textsuperscript{1}H and \textsuperscript{13}C spectra, where $B'$ was varied. The solid curves are the best-fit sums of two Lorentzian curves of equal amplitude but different sign, centered at $+1$ and $-1$ times the center field shown. %\rtext{increase font size where possible}
    }
    \label{fig:fig5}
\end{figure*}

Using the OPM to detect the longitudinal field component, the \textsuperscript{13}C magnetization in [1-\textsuperscript{13}C]-pyruvate can be observed during continuous parahydrogen bubbling by applying a short series of 180\si{\degree} toggling pulses and measuring the peak-to-peak signal amplitude during the inter-pulse windows as shown in \autoref{fig:fig5}a.  The figure-of-merit signal in this case clearly indicates the sample magnetization build-up and decay after the parahydrogen supply is turned on or off.  This is a trivial example for illustrative purposes. A clear future application would be for maximizing the sample magnetization by implementing on-the-fly choice of an optimized magnetic field sequence for hyperpolarization\cite{Eriksson2022JMR}, before the sample is removed from the magnetic shield.

Polarization may again be probed in situ in a more conventional style\cite{theis2011parahydrogen,theis2012zero,lee2019squid,blanchard2021towards,van2022relayed} by interrupting the sub-\si{\micro\tesla}-field polarization process and carrying out a Fourier-transform NMR experiment to quantify the amplitude of spin-specific resonances.  Spectra (\autoref{fig:fig5}b) recorded by the OPM at a \SI{4}{\micro\tesla} solenoid field provide high-resolution information on the heteronuclear-coupled spin states, revealing in-phase and anti-phase spin multiplets around the \textsuperscript{13}C and \textsuperscript{1}H Larmor frequencies, respectively.

For pyruvate we see the maximum \textsuperscript{13}C and \textsuperscript{1}H signal occurs at different fields. This results from the unequal proximity (and therefore unequal spin-spin \textit{J}-couplings) between the \textsuperscript{13}C/\textsuperscript{1}H spins and parahydrogen when transiently bound together in the catalytic intermediate. The overall field dependence can be fit by summing two Lorentzian curves with opposite amplitude, centered at $+1$ and $-1$ times the center frequency shown\cite{colell2017generalizing,blanchard2021towards}.  
It is the tilt of the OPM sensitive axis that enables a choice between these complementary approaches: the slower Fourier-transform approach senses changes to individual resonances through transverse magnetization detection, while longitudinal detection enables real-time tracking.

\section*{Discussion and Conclusion}
\label{sec:discussion}
In this work we have combined easily accessible magnetic field manipulations with sub-pT magnetometry to study nuclear-spin hyperpolarization with unprecedented (real-time) resolution.  This precision sampling approach appeals to many application areas, such as understanding the complex dynamics of magnetization, controlling and optimizing hyperpolarization in challenging or weakly controlled environments in real time, and in benchmarking quantum computations\cite{Seetharam2023SciAdvDigitalSimulation}.
Existing low-field polarizer instruments can easily be retrofitted with these measurement capabilities given the simple hardware needed, and, new implementations may be explored such as magnetometry during hyperpolarization in microfluidic systems\cite{Eills2022PNMRSsynergies}.  Real-time observation techniques may also be applied without shielding, using alternative magnetometers (e.g., free-induction decay OPMs, or giant-magnetoresistance sensors) that tolerate higher magnetic fields.

In the specific case of para-hydrogen-induced hyperpolarization, real-time detection unveils abundant information that is inaccessible by conventional NMR.  High-resolution, single-shot measurements can report on the differences between  
\textsuperscript{1}H and \textsuperscript{13}C state preparation under both naive and optimized adiabatic conversion schemes, and between different molecules, including subtle imperfections in experimental conditions.  
The advantage of direct vs.\ indirect (\textit{post-hoc}, point-by-point) sampling is taken to an extreme when we study magnetogenesis reversibility.  Here we see that magnetization profiles during successive adiabatic field cycles are not constant.  Such profiles also reveal direct signatures of distant dipolar fields (DDFs) that indicate -- without perturbing the system -- when the sample enters the high-magnetization regime.  In addition to this, the strengths of pulse-acquire NMR are not sacrificed: the OPM is capable of quantifying the DDF magnitude by measuring frequencies of off-axis magnetization precession. Finally, we observe in real-time chemical-exchange hyperpolarization in [1--\textsuperscript{13}C]-pyruvate, a widely used tracer in preclincal and clinical MRI studies of metabolic disease. 

We expect that the information provided by real-time observation techniques can facilitate fresh attempts to polarize specific molecules.  A candidate with high potential as a multi-pathway reporter of metabolic disease by MRI is \textsuperscript{13}C-labeled glucose\cite{Rodrigues2013NatureMedglucose}. Unlike pyruvate, this is significantly far from preclinical and clinical applications, in part due to the large number of spins that result in complex conditions for efficient hyperpolarization.

The magnetization tracking methods demonstrated here could also readily be extended from liquids, as shown, to hyperpolarized solids \cite{Tateishi2014PNAS,eichhorn2022hyperpolarized}
and to materials undergoing phase changes, such as melting or precipitation\cite{Knecht2021PNAS}. These are of interest because $T_1$ lifetimes in solids can be much longer, offering a haven for hyperpolarized magnetization storage. 
The sensitivity of the approach to molecular $J$-couplings, illustrated in \autoref{fig:fig2}, also opens the way to passive real-time tracking of chemical dynamics, including reactions and molecular conformational changes.

Besides use as a general method to study, machine-learn or optimize hyperpolarization, we expect the information provided by our approach to stimulate original ways to exploit hyperpolarized substances as sensors.  For instance, the reversible field-sweeping experiment of \autoref{fig:fig3} shows [1--\textsuperscript{13}C]-fumarate is a source of strong \textsuperscript{1}H/\textsuperscript{13}C magnetization that can be switched on and off repeatedly.  We speculate that this system could offer new opportunities for co-magnetometry (with simultaneous magnetic field sensing using the electrons in the OPM and the nuclear spins in the sample, or the co-located \textsuperscript{1}H and \textsuperscript{13}C nuclear spin species, which are strongly or weakly coupled depending on the magnetic field strength), or be envisioned as a toggleable polarized-nuclear target that is also lightweight and mobile enough to be used where current technologies do not reach, e.g., in space.

\section*{Materials and Methods}

\noindent \textbf{Parahydrogen-induced polarization:}
Hydrogen gas (H\textsubscript{2}) was prepared at $>$99\% para enrichment using a commercial generator (ARS Cryo) operating at 30\,K by flowing the gas over a hydrated iron (III) oxide catalyst (Merck KGaA, CAS 20344-49-4). The gas was stored at 10\,bar in an aluminium cylinder and used throughout the day for experiments.\\

\noindent \textbf{PHIP polarization of [1--\textsuperscript{13}C]-fumarate and [1--\textsuperscript{13}C]-maleate:}
To produce hyperpolarized [1--\textsuperscript{13}C]-fumarate, the precursor solution for all experiments was 125\,mM [1--\textsuperscript{13}C]-acetylenedicarboxylic acid disodium salt (Merck), 125\,mM sodium sulfite, and 3.5\,mM ruthenium catalyst (pentamethyl cyclopentadienyl tris(acetonitrile) ruthenium(II) hexafluorophosphate) (Merck, CAS 99604-67-8) in $\sim$99\% D\textsubscript{2}O, at pH 10-11.  For each measurement, 600\,µL of the solution was loaded by pipette into a 5\,mm screw-top NMR tube, and the tube was sealed. The screw top had two 1/16" (o.d.) PTFE capillaries passing through, acting as an inlet and an outlet for the hydrogen gas. The tube was pressurized to 5\,bar with parahydrogen, and its lower end placed into a mineral oil bath for the solution to warm (\SI{100}{\celsius}, 10\,s).  While still in the hot bath, parahydrogen was flowed through the tubes and bubbled through the solution (5\,bar for 60\,s). In most cases, the solution turned a pink/red color, indicating the reaction was near completion.  The tube was depressurized, opened, and the solution was extracted through a plastic capillary into a 1\,mL syringe. The solution was then injected by hand through a 1/16" (o.d.) PTFE capillary tube into a 2\,mL, 8-425 vial (Merck) located inside the shielded apparatus of \autoref{fig:fig1}b; the approximate delay between the end of the reaction and the start of field sweeps was 30\,s.

To produce hyperpolarized [1--\textsuperscript{13}C]-maleate, a similar procedure was used. The precursor solution was 125\,mM [1--\textsuperscript{13}C]-acetylene dicarboxylic acid and 5\,mM rhodium catalyst ([1,4-bis(diphenylphosphino) butane] (1,5-cyclooctadiene) rhodium(I) tetrafluoroborate) (Merck, CAS 
79255-71-3) in acetone-d\textsubscript{6}. 600\,µL of this solution was loaded into a 5\,mm screw-top NMR tube, and the same bubbling system was used for the reaction. Instead of using an oil bath, the tube was pressurized to 5\,bar and then heated for 10\,s with a heat gun until the solvent was boiling. Parahydrogen was then flowed through the sample for 45\,s under continual heating. After the reaction, the tube was cooled in water ($<$2\,s) and then depressurized and opened to extract the sample in the same manner as for fumarate. The sample was also injected into the magnetometer apparatus as described before.\\

\noindent \textbf{SABRE-SHEATH polarization of [1--\textsuperscript{13}C]-pyruvate:}
To hyperpolarize [1--\textsuperscript{13}C]-pyruvate using SABRE-SHEATH, a solution of 60\,mM [1--\textsuperscript{13}C] sodium pyruvate (Merck, CAS 87976-71-4), 6\,mM  IrIMes(COD)Cl and 40\,mM dimethyl sulfoxide-d\textsubscript{6} in methanol-d\textsubscript{4} was prepared. 1\,mL of this sample was loaded into a 2\,mL sample vial, the vial was sealed with a screw cap, and then placed into the shielded solenoid adjacent to the magnetometer. The sample vial cap featured a coaxial capillary (1/16" O.D.) with an inner glass capillary (0.15 mm o.d.) extending into the vial. These were used to bubble parahydrogen through the sample at 5\,bar. The catalyst was activated by 10\,mins of H\textsubscript{2} bubbling and the sample was cooled to approximately 10\,ºC by flowing ice-water around the sample vial.  The sample was used under these conditions for the reported experiments.\\

\noindent \textbf{Field control:} 
For magnetic field control and shielding, a multilayer MuMetal magnetic shield in a cylinder shape was used (Twinleaf model MS-1F, dimensions 30\,cm length and 25\,cm diameter).  Inside the shield, a solenoid coil (7.5\,mT/A, 15\,mm diameter) was positioned parallel to the cylinder axis.  Ramped magnetic fields were generated using a 12-bit digital-to-analog converter (DAC) integrated circuit (MCP4822, MicroChip Technology Inc.) whose output was connected to the solenoid.  The DAC was digitally interfaced with a microcontroller, provided as part of the ``NMRduino'' open-source magnetic resonance platform\cite{tayler2024nmrduino}.  The linear and constant-adiabaticity ramp profiles each comprised 800 equally spaced voltage points on a 1-\si{\micro\second}-precise time base.  Additional magnetic coils located inside the shield were used to adjust the background field at the magnetometer, with bias currents zeroed before the experiments, controlled by a low-noise controller (Twinleaf CSB-10).\\

\noindent\textbf{Optical magnetometry:} 
The OPM used for detection comprised a cube-shaped alkali-metal vapor cell as its sensitive element ($5 \times 5 \times 8$ mm$^3$), inside which \textsuperscript{87}Rb and N\textsubscript{2} buffer gas was contained and heated to \SI{150}{\celsius}.  A D1-resonant (\SI{795}{nm}), circularly polarized light beam was passed through one of the cube faces to optically pump the atomic spin polarization of the vapor.  A second, off-resonance beam was passed through one of the other faces, and out of the opposing side, to probe the atomic polarization via optical rotation.  

The remaining cube face was placed against the exterior of the solenoid coil, near to the NMR sample vial contained within.  A water cooling system was used to maintain a solenoid and vial temperature below \SI{30}{\celsius} (or \SI{10}{\celsius} in the case of the [1--\textsuperscript{13}C]-pyruvate experiments), which resulted in a standoff distance of 5-6 mm between the vapor cell and the vial.  Further details of the magnetometer can be found in past work\cite{Bodenstedt2021natcomm}.\\

\noindent\textbf{Data acquisition and processing:}
The rotation angle of the OPM probe beam was detected at a balanced polarimeter (Thorlabs PDB210A), which produced a differential output voltage linearly proportional to the $z$- and $x$-axis magnetic field amplitude, as illustrated in \autoref{fig:fig1}b.  The voltage was digitized by a second ``NMRduino''\cite{tayler2024nmrduino} device (16 bit, $\sim$\SI{150}{\micro\volt\per bit}, 1--5 kHz sampling rate) for a few seconds to minutes, depending on the experiment.  The signal was streamed to a computer via USB and displayed on-screen in real time after digital low-pass filtering.  The raw signal data were also stored in a file on the computer for future retrieval, processing and plotting.  Data processing operations to generate the plots shown in \autoref{fig:fig1}--\autoref{fig:fig5} involved simple operations in Mathematica (Wolfram Inc.) including: (i) low-pass Hamming or moving-average filters to suppress 50 Hz mains-electricity noise, which was the dominant noise source of the magnetometer, and (ii) Fourier transformation, as in \autoref{fig:fig4}.  A record of these operations is contained in the Supporting Information.\\

\noindent\textbf{Simulations:}
All simulations were carried out using the SpinDynamica packages for Mathematica.\cite{bengs2018spindynamica} All simulations involved three spins, two protons and a carbon-13, and the interactions considered were the Zeeman interaction between the spins, the applied magnetic fields (in $x$, $y$, or $z$) and the spin-spin $J$\nobreakdash-couplings (untruncated couplings in all cases). Chemical shifts were omitted because at such low fields these terms are negligible. Relaxation was included phenomenologically for the simulations in \autoref{fig:fig1}c and \autoref{fig:fig3}, as a constant rate of signal decay over time, independent of field or the type of spin order.

The density operator of the fumarate/maleate systems immediately after the PHIP process was assumed to be pure nuclear-spin singlet order (of the two protons $I_1$ and $I_2$) secularized under the $J$-coupling Hamiltonian plus a 50\,$\mu$T background field to approximate Earth's field. 
The $J$\nobreakdash-couplings used for the simulations are shown in \autoref{fig:fig1}c, and these couplings were modified within the given errors (50\,mHz) for each simulation to fit to the data.
The field profiles that are plotted in \autoref{fig:fig1}c, \autoref{fig:fig2} and \autoref{fig:fig3} were obtained by propagating this initial density operator under the applied field sweep(s), and evaluating the projection of the sample's magnetization along a vector tilted 30º from the $z$ axis towards the $y$ axis, at each time point. This is given by the observable:
\begin{equation}
    \text{signal}\text{(t)} \propto \text{cos}(30^{\circ}) \langle M_z \rangle \text{(t)} + \text{sin}(30^{\circ}) \langle M_y \rangle \text{(t)},
\end{equation}
where the operator for the total magnetization along the axis $\xi = x, y, z$ is: $M_\xi = \gamma_\text{I} I_{1\xi} +\gamma_\text{I} I_{2\xi} +\gamma_\text{S} S_{3\xi}$. The 30º rotation corresponds to the magnetometer's sensitive axis.

A Mathematica notebook is included in the Supporting Information containing the raw simulation code with comments.

\section*{Data, Materials, and Software Availability}
Experimental data files supporting the findings of this study plus a Mathematica notebook containing processing and simulation code with comments are freely available via the OPENAIRE database, hosted on Zenodo: URL \url{https://doi.org/10.5281/zenodo.13752103}. All other data are included in the manuscript and/or \href{https://www.pnas.org/lookup/doi/10.1073/pnas.2410209121\#supplementary-materials}{supporting
information}.

\section*{Acknowledgments}
The work described is funded by: 
the Spanish Ministry of Science MCIN with funding from European Union NextGenerationEU (PRTR-C17.I1) and by Generalitat de Catalunya ``Severo Ochoa'' Center of Excellence CEX2019-000910-S; 
the Spanish Ministry of Science projects SAPONARIA (PID2021-123813NB-I00), MARICHAS (PID2021-126059OA-I00), SEE-13-MRI (CPP2022-009771) plus RYC2020-029099-I and RYC2022-035450-I, funded by MCIN/AEI /10.13039/501100011033; 
Generalitat de Catalunya through the CERCA program;  
Ag\`{e}ncia de Gesti\'{o} d'Ajuts Universitaris i de Recerca Grant Nos. 2017-SGR-1354 and 2021 FI\_B\_01039; 
Fundaci\'{o} Privada Cellex; 
Fundaci\'{o} Mir-Puig;
and the BIST--“la Caixa” initiative in Chemical Biology (CHEMBIO); the Helmholtz Association (DB002399).
The project has received funding from the European Union’s Horizon 2020 Research and Innovation Programme under the Marie Sk\l{}odowska-Curie Grant Agreement 101063517.
We are also grateful to Andreas Trabesinger and Sven Bodenstedt for discussions.

\section*{References}
\bibliography{references}

\end{document}